\documentclass{quick-article}
\usepackage[utf8]{inputenc}

\usepackage[normalem]{ulem}  % strike-out
\usepackage{mathtools}   % cramped etc.

\def\algPC{\textit{Probabilistic Counting}}
\def\algAS{\textit{Adaptive Sampling}}
\def\algAC{\textit{Approximate Counting}}

\def\algLL{\textsc{LogLog}}

\def\algHLL{\textsc{HyperLogLog}}
\def\etal{\textit{et al.}}

\title{The Story of \textsc{HyperLogLog}:\\How Flajolet Processed Streams with Coin Flips}
\author{J\'{e}r\'{e}mie O. Lumbroso\thanks{J\'{e}r\'{e}mie O. Lumbroso Hall, Department~of~Computer~Science, Princeton
    University, 35~Olden~Street, Princeton, NJ 08540, USA.
    \email{lumbroso@cs.princeton.edu}}}
\date{December 2013}

\begin{document}

\maketitle

\begin{abstract}
    This article is a historical introduction to data streaming algorithms that was written as a companion piece to the talk ``\emph{How Philippe Flipped Coins to Count Data}'', given on December 16th, 2011, in the context of the conference in honor of \emph{Philippe Flajolet and Analytic Combinatorics}~\cite{BibPFAC}. The narrative was pieced together through conversations with Philippe Flajolet during my PhD thesis under his supervision, as well as several conversations with collaborators after his death. In particular, I am deeply indebted to Nigel Martin for his archival records. This article is intended to serve as an introductory text presenting Flajolet's data streaming articles in a projected set of complete works.
\end{abstract}

\vspace*{0.5truecm}

\begin{small}
  \begin{flushright}\textit{
    As I said over the phone, I started working on your algorithm when\\
    Kyu-Young Whang considered implementing it and wanted
    explanations/estimations.\\
    I find it simple, elegant, \sout{surprisingly}
    amazingly powerful.}\\[2mm] 
  --- \textsc{Philippe Flajolet}, in a letter to G. Nigel N. Martin
  (1981).
  \end{flushright}
\end{small}

\bigskip

\noindent By 1981, when he first encountered the seeds of what was going
to become streaming algorithms at IBM San Jose, Philippe Flajolet had
already distillated an impressive set of tools for the analysis of
algorithms. But most of these techniques were more or less developped for
the problems they were supposed to help solve, and Flajolet was interested
in finding completely unrelated problems that could be approached using
the same techniques.

Probabilistic streaming algorithms, which Nigel Martin and Philippe
Flajolet pioneered, proved an exciting such topic. Far from having been a
passing interest, Flajolet repeatedly returned to them over more than two
decades. His contributions to this subject have been significant but also
serve to illustrate a different aspect of his research interests: although
these results were eminently mathematical, they showed his understanding
of, and appreciation for, implementation level details.

And as this chapter contains a survey by Flajolet himself~\cite{PF180},
which already goes a long way exposing the mathematical concepts involved
in these algorithms, we have seized the opportunity to approach this topic
from a rather more historical perspective.

\section{Approximate Counting (1982)\label{sec:morris}}

As a starter, we look at an algorithm Flajolet first wrote about in
1982~\cite{RR0153}. This algorithm is different from the others which will
be discussed in this chapter, most notably in that it does not require
hash functions. Instead, it is a conceptually simpler introduction to the
concept that some theoretical bounds---here the information-theoretical
limit that $\log_2 n$ bits are needed to count up to $n$---can be
circumvented by making \emph{approximations} using \emph{probabilistic}
tools.

\subsection{Context: spellchecking with no dictionary?}

The researchers developing Unix at Bell Labs in the mid 70s were
fascinated by text processing. Robert Morris wanted to count the
occurrences of \emph{trigrams} in texts---overlapping substrings of three
letters. These counts could then be used by \textsf{typo}, a
statistic-based spellchecker included in early UNIX distributions, at a
time where dictionary-based approaches were out of the question for
storage (size and access speed) reasons, see~\cite{MoCh75}
and~\cite[\S 3.2]{McChMo78}.

Unfortunately in this pre-16-bit era, Morris could only fit
$\cramped{26^3}$ 8-bit counters into the memory of his PDP-11 mainframe,
thus limiting the maximum count to 255: much too small a range to gather
any sort of useful trigram count.

Thus instead of maintaining exact counters, Morris suggested making
increments in a probabilistic manner. But quickly pointed out that doing
so using constant probabilities is not very useful: either the probability
of an increment is too large, and the reach is not significantly improved
(for example, if you increment every other time, that is with probability
$1/2$, then you only allow yourself to count up to 511: you only spare one
bit, and the tradeoff is a 50\% error); or the probability of an increment
is too small, and thus the granularity is too large, in particular making
small counts consistently over-estimated (for instance, with a probability
of $1/25$, you cannot keep track of values smaller than $25$). This
approach is also discussed as ``\emph{direct sampling}'' by Flajolet at
the end of his article.

This suggests the probability of making an increment should not be
constant, but instead depend on the current value of the counter. In
essence, Morris' idea~\cite{Morris77} is that, with a probability of
increment exponential in the value of the counter, it is possible to keep
track not of the number $n$ to be counted, but of its logarithm,
significantly saving bits (in fact, Morris and his colleagues called the
algorithm \emph{logarithmic counter})\footnote{Furthermore, this idea is
  related to unbounded search in an ordered table, and in recent times has
  often been presented as such: you are looking for an entry $x$ in an
  ordered table of unknown and/or infinite size, so you first find out in
  which geometric interval $[2^k, 2^{k+1}[$, $k \geqslant 0$, $x$ is, then
  proceed to do dichotomic search in this interval (the way the intervals
  are subdivided impacts the complexity, see~\cite{BeYa76}).}.

\subsection{Algorithm}

The formulation everybody is familiar with, as well as the name {\algAC},
are due to Flajolet, who, in so doing, contributed greatly to the overall
popularity of the algorithm\footnote{An overwhelming majority of citations
  to Morris' original article date from after 1985, and were usually made
  in tandem with Flajolet's paper.}.

%% BriTODO: mal dit "Let N be the value to be counted to" -> tried

Let $N$ be the value we would like to keep track of, i.e., the number of
calls to \textsc{AC-AddOne}; and let $C$ be the value of the (approximate)
counter, initially
%\footnote{The counters are in truth set to
%  $0$, but the increment from $0$ to $1$ is done deterministically.}
set to $1$. If $\BernoulliLaw{p}$ denotes a \emph{Bernoulli random
  variable} (colloquially known as a biased coin flip), equal to $1$ with
probability $p$ and $0$ with probability $1-p$, then adding to and
retrieving the value of the counter is done, in its most basic version,
with the following procedures:
\begin{align}
  \begin{array}{l}
    \mbox{\textsc{AC-AddOne}$(C)$:}\\[-0.4em]
    \quad C \leftarrow C + \BernoulliLaw{\cramped{1/2^C}}
  \end{array}\qquad\qquad
  \begin{array}{l}
    \mbox{\textsc{AC-Estimate}$(C)$:}\\[-0.4em]
    \quad \mbox{\textbf{return} }\cramped{2^C}-2
  \end{array}
\end{align}
to the effect that at all times, an estimate of $N$ is given by
$N\approx\cramped{2^C}-2$. Indeed when the counter $C$ is equal to $1$,
the probability of making an increment is $1/2$, thus it will take on
average $2$ calls to \textsc{AC-AddOne} for the counter $C$ to go from $1$
to $2$; it then takes $4$ calls on average to go from $2$ to $3$; and more
generally, $\cramped{2^k}$ calls to go from $k$ to $k+1$, to the extent
that it requires (on average)
\begin{align}
  2^1+2^2+\ldots+2^k = \sum_{i=1}^k 2^i = 2^{k+1}-2
\end{align}
calls to \textsc{AC-AddOne} for the counter $C$, initially set to $1$, to
be equal to $k+1$.

The accuracy of such a scheme is of roughly one binary order of
magnitude---which can be deduced from elementary observations. This
accuracy can be improved by changing the base of the logarithm, and making
probabilistic increments with probability $\cramped{q^{-C}}$ instead of
$\cramped{2^{-C}}$, in which case the estimator then becomes
\begin{align}
  f(C) := \frac{q^{C} - q}{q-1}
\end{align}
such that the expected value of $f(C)$ after $N$ increments is equal to
$N$. The counter will then perform within one $q$-ary order of magnitude;
if $q\in(1,2)$ the accuracy is expected to be improved over the binary
version, with a space tradeoff.

While Flajolet greatly clarified Morris' original algorithms, his other
main contribution is to have analyzed them with great finesse. He obtained
a more precise characterization of the expected value and of the accuracy
involving periodic fluctuations. To this end, he studied an harmonic sum
expressing the expected value of $C$ using the \emph{Mellin transform}
discussed in more detail in Chapter~4 of Volume~III. It is worthwhile to
note that Flajolet was particularly excited to find, first in {\algPC},
and then (also through Martin~\cite{HeLaMaTo82}) {\algAC}---the
analysis of both involving such a complex harmonic sum, or in his words:
``\emph{I completed the analysis of {\algAC} and (again!) it has a fairly
  interesting mathematical structure}'' (1981). The results provided by
Theorem~2 or Section~5, with an expression given as the sum of
a linear/logarithmic term, a precise constant term and a trigonometrical
polynomial, typically exemplify the sort of fascinating sharp yet easy
results yielded by Mellin analysis.

\subsection{Recent extensions and applications}

In addition to the statistical application introduced as motivation,
{\algAC} has been used recurrently in a number of different data
compression schemes, where many frequency statistics must be collected,
but where their absolute accuracy is not critical (see for
instance~\cite{HeLaMaTo82}, through which Philippe initially discovered
the algorithm, or~\cite[\S 3.1]{McChMo78}). But although these
applications highlight the space-saving aspect of {\algAC}, it would be
mistaken to think that {\algAC} is no longer relevant, with nowadays' huge
storage sizes.

The algorithm has reached great recognition in the streaming literature,
as it efficiently computes $\cramped{F_1}$, the first frequency moment
(in the terminology of Alon {\etal}~\cite{AlMaSz96}); it is thus often
cited for this reason. Beyond that, it has been extended and used in a
number of interesting, practical ways. Here are several recent examples.

In 2010, Mikl\'{o}s Cs{\H{u}}r{\"o}s introduced a floating-point version
of the counter~\cite{Csuros10}, where accuracy is set in a different way:
instead of picking a base $q$ for the logarithmic count, Cs{\H{u}}r{\"o}s
suggests splitting the counter into a $d$-bit significant and a binary
exponent. The total bits used to count up to $N$ is $d+\log\log N$ bits,
but the appreciable advantage is that small counts, up to $N=2^d-1$, are
exact\footnote{Another advantage is that the algorithm only requires
  random bits as a source of randomness---instead of random real
  values---and requires only integer arithmetic, making efficient
  implementations easy.}. This variant was developed in the context of
mining patterns in the genome~\cite{CsNoKu07}, and coincidentally uses an
approach which is reminiscent of Morris' original application within the
\textsf{typo} program.
%%\footnote{It must be noted that this original
%%  application of {\algAC} is very scarely known throughout the literature,
%%  and it is unlikely that it was the source of inspiration of any current
%%  work on data-mining.}.

In 2011, Jacek Cich\'{o}n and Wojciech Macyna~\cite{CiMa11}, in another
ingenious application, suggested\footnote{Using {\algAC} in the context of
  flash memory had already been suggested independently by
  \cite{ZuBaTo09}, but only as an off-hand comment.} using {\algAC} to
maintain counters keeping track of the way data blocks are written in
flash memory. Indeed, flash memory is a flexible storage medium, with one
important limitation: data blocks can only be written to a relatively
small number of times (this can typically be as low as 10\,000 times). As
a consequence, it is important to spread out data block usage; this is
most routinely done by tracking where data has been written through
counters stored on a small portion of the flash memory itself. Cich\'{o}n
and Macyna point out that using {\algAC} in this context is pertinent not
only because it cuts down on the storage of the counters, but also because
the \emph{probabilistic increment} decreases the number of times the
counters are actually modified on the flash memory.

This perfectly illustrates the fact that the probabilistic increment at
the heart of {\algAC} can be used for two very different reasons: either
when \emph{storing} the increments is costly; or when the \emph{action of
  incrementing itself} is costly. As a parting note, here is another
illustration: suppose you had a counter stored remotely; each increment
would require some communication complexity (the size of the message sent
remotely to increment the counter); this communication complexity could be
considerably decreased, from $O(N)$ to $O(\log N)$, if an {\algAC} type
idea were used.

\section{An Aside on Hash Functions\label{sec:hash}}

With the exception of his paper on {\algAC} which we have just covered,
the remainder of Flajolet's work on probabilistic streaming algorithms
uses, at its core, \emph{hash functions}.

\subsection{Back in the day.}

%% BriTODO: un rappel un peu plus précis ? -> tried

Hash functions (initially also referred to as \emph{scatter storage}
techniques) were created in the 1950s for the generic storage/retrieval
problem, as an alternate method to, for instance, sorted tables and binary
trees~\cite[pp. 506-542]{Knuth73}. The abstract premise is that instead of
organizing records relative to each other through various schemes of
\emph{comparisons}, the position of a record $x$ in a table is calculated
directly by applying a hash function as $h(x)$. As a consequence, with
care, hash tables are robust data structures which can have storing/access
times that are independent of the number of items stored, and have become
extremely popular. In additionally hash functions have found a number of
unrelated uses (fingerprinting, dimensionality reduction, etc.).

It is plain to see that the issue here is \emph{collision}, that is when
two different elements $x\not=y$ map to the same value $h(x) = h(y)$. At
first, hash functions were very specifically designed (as Knuth says, like
a ``puzzle'') for a particular set of well defined elements, so as to
scrupulously avoid any collisions. Predictably that approach was too
unflexible and complex, and soon the goal was only to design hash
functions that spread the data throughout the table to attenuate the
number of collisions\footnote{Another reason why it was important to
  spread out the load was that \emph{linear probing}---where upon
  a collision an element is placed in the closest empty spot---was
  a popular method to resolve collisions; if elements are clustered
  together then the next empty spot is much further away from the initial
  hash index.}. These properties naturally had to be formalized so
algorithms using hash functions could be analyzed.

%%% NOTE: ces trois paragraphes sont très mauvais et à réécrire.

Thus hash functions began being modelled as associating \emph{uniform
  random variables} to each record. At first, this model~\cite{Peterson57}
was very much an idealized approximation. But eventually, as somewhat of
an unintended side-effect, hash functions ended up actually becoming good
at \emph{randomizing} data: turning any sort of data into some
pseudo-uniform data. Eventually, algorithms began taking advantage of this
probabilistic aspect; one particular notable example is Bloom
filters~\cite{Bloom70}\footnote{The name seems to have been coined in
  a 1976 paper by Severance and Lohman~\cite{SeLo76}.}, which basically
introduced the paradigm of ``\emph{[advocating] the acceptance of
  a computer processing system which produces incorrect results in a small
  proportion of cases while operating generally much more efficiently than
  any error-free method''}~\cite{JaPa73}.

Experiment simulations suggested this approach worked surprisingly well,
and this usage was cemented in 1977, when Carter and Wegman~\cite{CaWe77,
  CaWe79} showed how to build hash function following increasingly
stringent probabilistic requirements---including uniformity---therefore
providing solid theoretical ground by which to justify the practice.

Yet Carter and Wegman's ``universal hash functions'' were rarely used in
practice on account of their computational inefficiency, and simpler hash
functions yielded surprisingly good results. Quite recently, Mitzenmacher
and Vadhan~\cite{MiVa08} discovered that the reason for this success is
that even simple hash functions are very efficient at exploiting the
entropy of the data.

%% BruTODO: Long. Pas indispensable.

\subsection{From data to uniform variables: reproducible randomness.}

Let $\mathcal{U}$ be the possibly infinite set (or \emph{universe}) of
elements that can be hashed; a hash function can be modeled theoretically
by a function $h:\mathcal{U}\rightarrow \cramped{\{0,1\}^\infty}$ which is
said to \emph{uniformize} data, when it associates to every element an
infinite sequence of random bits, or Bernoulli variables of parameter
$p=1/2$, that is
\begin{align}
  \forall x\in\mathcal{U},\quad h(x) = y_0\,y_1\,y_2\,\cdots \quad
  \text{ such that } \quad
  \forall k\in\N, \quad \prob{y_k=1}=\frac{1}{2}\text{.}
\end{align}
(This definition differs from the more traditional one which has hash
functions output an integer, but these two definitions are equivalent and
related by binary expansion.)

Of crucial importance is the apparent contradiction that the hash functions
are, by nature, \emph{functions}---thus a given hash function $h$ always
associates to an element $x$ the same value $h(x)$---while providing the
illusion of randomness. In a strong sense, hash functions provide
\emph{reproducible randomness}, and this concept is at the heart of many
probabilistic streaming algorithms.

\section{Probabilistic Counting (1981-1985)\label{sec:pc}}

This \emph{Probabilistic Counting} algorithm, as all further ones to be
discussed in this introduction, is concerned with efficiently
approximating the \emph{number of distinct elements} (also called
\emph{cardinality}) in a stream, which may of course contain repetitions.

%% BriTODO: ? BruTODO: Bof.

Contrasting with a common, unfortunately lasting,
misconception~\cite{AlMaSz96}, the genesis of \emph{Probabilistic
  Counting} was thoroughly practical, to the extent that versions of the
algorithm were implemented and \emph{in production}~\cite{AsScWh87} well
before the algorithm was fully analyzed. This makes the contribution
unlike most of the litterature, essentially theoretical in nature (such as
Alon~\etal~\cite{AlMaSz96} or more recently Kane~\etal~\cite{KaNeWo10}),
since then published on data streaming algorithms.

\subsection{Historical context: the birth of relational databases}

In the early days of \emph{database management systems}, at the end of the
60s, accessing data required an intricate knowledge of how it was stored;
queries needed to be hard-coded by programmers intimately familiar both
with the system and with the structure of the database being queried. As
a result, databases were both unwieldy and costly.

Eventually, following the ideas of Edgar Codd at IBM in the
70s~\cite{Codd70}, there was a large push towards \emph{relational
  databases} that could be designed and queried through a high-level
language. Obviously, a crucial concern was \emph{query
  optimization}---ensuring that the computer-effected compilation of these
high-level queries into low-level instructions, produced results within
the same order of efficiency as the human-coded access routines of yore.
And it soon became apparent the number of distinct elements (in a data
column) was the most important statistic on which to base optimization
decisions, see~\cite{SeAsChLoPr79} or~\cite[p. 112]{SystemR76}.

Martin was an IBM engineer in the UK, who worked on one of the first
relational databases~\cite{Todd76}. When the project came to term in 1978,
Martin was granted a sabbatical to conduct original research at the
University of Warwick, during which he published works on extendible
hashing~\cite{Martin79,ChKn94} and data
compression~\cite{Martin79b,HeLaMaTo82}. Eventually, he was called to IBM
San Jose, to present his unpublished ideas; one of which---influenced by
his work on hashing, the emerging ideas on approximating searching and
sorting\footnote{Indeed, Mike Paterson~\cite{MuPa80} was at Warwick at the
  time, and a close colleague of Martin.} and his prior knowledge of
databases---was the original version of \emph{Probabilistic Counting}.

%%%% NOTE: no mod.

\subsection{Core algorithm\label{subs:pc-core}}

We assume we have a hash function $h$, which transforms every element
$\cramped{y_i}$ of the input data stream, into an
infinite\footnote{Working with \emph{infinite} words is a theoretical
  commodity: of course, in practice, the words are of fixed size---32 or
  64 bits usually---, and a precise discussion on this is included in the
  paper. The bottom line is that this in no way limits the algorithm.}
binary word $h(\cramped{y_i})\in \set{0,1}^\infty$, where each bit is
independently $0$ or $1$ with probability $1/2$. The algorithm is based on
the \emph{frequency of apparition of prefixes} in these random binary
words. Specifically, they were interested in the position of the leftmost
$1$.

Since each bit is independently $0$ or $1$ with probability $1/2$, we
expect that: one in every two words begins with $1$; one in every four
words begins with $01$; one in every $\cramped{2^k}$ word begins with
$\cramped{0^{k-1}1}$. Conversely, it is reasonable to assume that in
general if we see the prefix $\cramped{0^{k-1}}1$ which occurs with
probability $\cramped{1/2^k}$, we can assume there are about
$\cramped{2^k}$ words in total.

The algorithm keeps track of the prefix it has seen by maintaining a
\emph{bitmap} or vector of bits, initially all set to $0$, bit $i$ is set
to $1$ when a prefix of length $i+1$ has been seen. It then make an
estimate based on the \emph{position of the leftmost zero in this bitmap},
which we note $R$.

\paragraph{Example.}
Consider the following stream $S$---in which the infinite words have been
truncated to 5 bits, repetitions have been removed, and the prefixes we
are interested in have been bolded,
\begin{align*}
  S=\textbf{1}0000,\textbf{1}1101,\textbf{00001},\textbf{1}1011,
  \textbf{01}100,\textbf{1}0110,\textbf{1}0111,\textbf{001}11
\end{align*}
Once the stream has been processed, the bitmap is equal to 11101; the
position of the leftmost zero (remember positions start in $0$) is $3$. We
can thus make our guess that there are about $\cramped{2^3}$ distinct
elements in stream $S$.

Had $S$ contained repetitions, the final value of the bitmap (and
consequently our estimate) would have been the same. This is because we
are \emph{projecting} the rank of the leftmost one onto the bitmap---and
projections are insensitive to repetitions.

\subsection{Analysis: no algorithm without math.\label{subs:nomathnoalg}}

So let $R$ be the position of the leftmost zero in the bitmap. Though by
construction, it is reasonable to consider that this random variable is on
average close to $\log_2 n$, in truth, $R$ has a systematic \emph{bias} in
the sense that there is some $\phi$ such that $\cramped{\expect[n]{R}
  \approx \log_2(\phi n)}$. As a consequence, if we simply take
$\cramped{2^R}$ as an estimate of the number $n$ of distinct elements,
then we will be off by non-negligible fraction.

%%%\begin{figure}
%%%  \centering
%%%  \includegraphics[scale=0.6]{../images/gnn-final-tweak.pdf}
%%%  \caption{\label{fig:orig-tweak}In the original algorithm, as designed by
%%%    Martin, the observed bias was attenuated using an ad-hoc
%%%    correction based on observing outlier bits.}
%%%\end{figure}

%% Remark (Kilian): Peut-être, quand tu emets un jugement, le faire de
%% façon plus douce.

Martin had noticed this, and introduced some ad-hoc correction: look
at the three bits following the leftmost zero; depending on their value,
adjust $R$ by $\pm1$
%%%(the original type-up detailing this point is shown
%%%%in Figure~\ref{fig:orig-tweak}).
While the reasoning behind this correction was clever, and it \emph{does}
somewhat concentrate the estimates while decreasing the bias, it does not
remove it: the estimates produced remain significantly biased.

In essence, this algorithm is in the uncommon position of \emph{requiring}
complex mathematical analysis within its design for its correctness---not
just its complexity analysis. This situation would be aptly described by
Flajolet's creed, ``\emph{no math, no algorithm}''; and one of the main
results of the paper~\cite[Theorem~3.A]{PF050} was to determine that the
expected value of the statistic $R$ is
\begin{align}
  \expect[n]{R} = \log_2(\phi n) + P(\cramped{\log_2}n)+o(1)\text{,}
\end{align}
where $P$ is an oscillating function of negligible amplitude, so that
indeed we may consider $\cramped{2^R/\phi}$ an \emph{unbiased} estimator
of the number of distinct elements.

\paragraph{Fascinating constants.}
Before we move onto how to make this algorithm useful in practice, I wish
to make a small digression and discuss this correction constant. The
constant $\phi\approx 0.77351\ldots$ is given exactly by
\begin{align}\label{eq:pcconst}
  \phi = 2^{-1/2}\smexp{\gamma}\frac 2 3 \prod_{p=1}^\infty
  \left[\frac{(4p+1)(4p+2)}{(4p)(4p+3)}\right]^{(-1)^{\nu(p)}}
\end{align}
where $\gamma$ is Euler's gamma constant and $\nu(p)$ is the number of
$1$-bits in the binary representation of $p$. %Jean-Paul
Allouche noticed that this constant was related to an unexpected identity
due to %Jeffrey
Shallit~\cite[\S 5.2]{AlSh98}, which provided the starting point for a
simplification. Using mainly the identity
\begin{align}
  \prod_{k=2p}^{2p+1} \left[\frac{2k+1}{2k}\right]^{(-1)^{\nu(p)}} = 
  \left[\frac{(4p+1)(4p+2)}{(4p)(4p+3)}\right]^{(-1)^{\nu(2p)}}
\end{align}
we can obtain the (much slowly converging) expression
\begin{align}
  \phi = \frac{\smexp{\gamma}}{\sqrt{2}} \prod_{p=1}^\infty
  \left[\frac{2p+1}{2p}\right]^{(-1)^{\nu(p)}}\text{.}
\end{align}
Some additional details are provided by Steven Finch in his book on
mathematical constants~\cite[\S 6.8.1]{Finch03}.

What is particularly notable is that the elegance and specificity of this
constant is the result of Flajolet's ``hands-on'' analysis, based on the
\emph{inclusion-exclusion principle}, which is where the number $\nu(p)$
of 1-bits in the binary representation of $p$ comes from. Indeed, the
Mellin transform of the probability distribution of $R$ contains the
Dirichlet function associated with $\nu(p)$
\begin{align}
  N(s) = \sum_{k=1}^\infty \frac{(-1)^{\nu(k)}}{k^s}\text{.}
\end{align}
The product in \eqref{eq:pcconst} results from grouping the terms in this
Dirichlet function by four. Although the tools Flajolet has developed
since would allow for a much simpler and straightforward analysis, these
would generally not yield such closed-form expressions.

Interestingly, Kirschenhofer, Prodinger and Szpankowski first published in
1992 an alternate analysis of the main estimator~\cite{KiPrSz92, KiPrSz96}
which illustrates this well. Instead of using the inclusion-exclusion
principle, they frame the analysis of the algorithm in terms of splitting
process, which Flajolet had partially written about some years
before~\cite{PF034}. Let $R$ be the statistic used by {\algPC} (the
leftmost zero in the bitmap) which we have described before, its probability
generating function can be described recursively
\begin{align}
  F_n(u) = \expect[n]{u^R}\qquad\text{and}\qquad
  F_n(u) = \frac 1 {2^n} + u\sum_{k=1}^n \binom{n}{k} \frac 1 {2^n} F_k(u)\text{.}
\end{align}
To obtain this recursion, we consider the bit-vector of all $n$ hashed
values, bit after bit, as though they were iterations. On the first
iteration, the probability that all first bits are $1$ is
$\cramped{1/2^n}$, and thus the rank of the leftmost zero in the bitmap
will be $0$---this contributes $1/2^n$ to the term $u^0$; or else, there
is at least one hash value of which the first bit is equal to $0$, and
thus we make a recursive call with $u$ as multiplicative factor.

Once this functional equation is obtained, the subsequent steps are (now)
standard, as we will see: iteration, Poissonization, Mellin. This type of
analysis is very similar to that of {\algAS} (see Section~\ref{sec:as}),
and reflects how our angle of approach has evolved since Flajolet's
initial analysis of {\algPC}. The corrective constant which the authors
find is
\begin{align}
  \log_2 \xi = -1-\frac 1 {(\log 2)^2} \int_{0}^\infty \smexp{-x}
  \prod_{j=0}^\infty\left(1-\smexp{-x2^{j+1}}\right)\frac {\log x} x \drm x
\end{align}
and is expected to satisfy $\xi = \phi$. A direct proof can be derived (as
shown by Allouche), and indeed, through numerical integration, we find
$\xi\approx 0.77351\ldots$ in good agreement with Flajolet's calculations.

\subsection{Towards an effective algorithm}

Although the algorithm, at this point, is unbiased, the estimates are
typically dispersed by one binary order of magnitude---as expected from
the fact that $R$ can only take integer values.

To improve the accuracy, we could simply run $m$ simultaneous instances of
the algorithm on the same stream, but using a different random hash
function for each instance; if we then average these $m$ estimates, the
central limit theorem states this would increase the accuracy by a factor
of $1/\sqrt{m}$.

This method, however, is not desirable for several reasons: even assuming
we were able to obtain $m$ good independent uniform hash functions, the
computational cost would be huge, especially in light of the fact that so
few of hashed values are actually useful\footnote{If the stream has $N$
  total elements, $n$ of which are distinct, then---per a classical result
  on \emph{records} in permutations---only about $O(\log n)$ of these
  values are expected to change the state of the bitmap; the rest are just
  ignored.}.
 
\paragraph{Stochastic averaging: making the most out of a single hash function.}
The \emph{stochastic averaging} technique simulates running many
concurrent versions of the algorithm using different hash functions, while
only using one single hash function---thus at a fraction of the
computational cost. As a tradeoff, it delays the asymptotic regime for
well-understood reasons, and introduces non-linear distortions.

Instead of running the algorithm in parallel with several hash functions,
then taking the average, a very similar effect can be reproduced by
splitting the main stream into several substreams. This is done by
sampling the first few bits of the hash value to determine in which stream
place the value, and discarding these bits. The averaging is called
\emph{stochastic} because every instance of an element is distributed to
the same substream (instead of just randomly distributing all occurrences
in the substreams, which would be useless, as the cardinality of
a substream would have no relation with the cardinality of the whole).

\begin{figure}[h]
  \centering
  \includegraphics[scale=0.8]{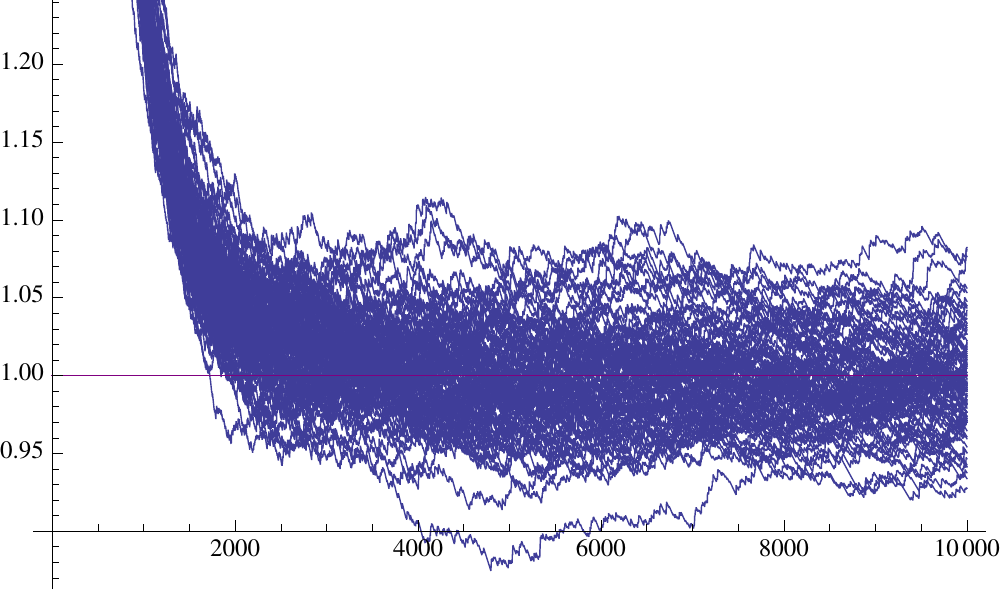}
  \caption{\label{fig:pcsa-distort}This plot represents the evolution of
    the accuracy (ratio of the estimate to the actual cardinality) as
    a function of the actual cardinality, during 100 concurrent runs of
    {\algPC}, on a stream containing $n=10\,000$ distinct elements and
    split into $m=512$ substreams. The estimates are seriously distorted,
    up until about $n=6m$, that is $3072$.}
\end{figure}

One undesirable side-effect of this technique is that the asymptotic
regime is significantly delayed, as shown in
Figure~\ref{fig:pcsa-distort}. Indeed while the original algorithm
provides comparatively accurate estimates throughout its whole range, we
now split the stream into $m$ substreams---and the quality of the
resulting estimates depends intricately on \emph{how many} substreams
actually contain elements. It is plain to see that if $n \ll m$ then the
problems are compounded: most substreams will be empty; those that aren't
will only contain a small fraction of the values. As a result, the final
average would be significantly worse than what would have been obtained
without stochastic averaging.

Empirical observations suggest that these distortions can be ignored for
$n>6m$, although recent work shows that for smaller values of $n$ the
distortions can be corrected~\cite[\S 3]{Lumbroso10}. The original paper
suggested keeping exact counts up to a certain threshold, and then
switching to {\algPC}; we will see in next section a different estimation
algorithm, {\algAS}, that does not have this issue with small
cardinalities, and also how Philippe Flajolet and Marianne Durand found an
elegant alternative solution when designing {\algLL}.

%%%% END NOTE

\section{Adaptive Sampling (1989)\label{sec:as}}

In 1984, Mark Wegman---of the universal hash function fame---suggested, in
private communications, a new algorithm for cardinality estimation, which
he named \emph{Sample counting} and which avoided the problem of
distortions of \emph{Probabilistic Counting} for small cardinalities.

\paragraph{Description.}
Wegman's algorithm uses the uniformizing properties of hash functions, as
described in Section~\ref{sec:hash}, to construct a subset containing
a \emph{known proportion} of all elements in the data stream.

It does so \emph{adaptively}: it initially assumes all elements in the
stream will fit into a cache with $m$ slots; then as it gets evidence to
the contrary (because the cache overflows), it decides to only keep 50\%
of all objets, and if proven false again then 25\%, and so on. And
finally, the selection of a subset of elements is done by restricting the
hash value of objects that can be in the cache: for instance, if the only
elements allowed in the cache are those with hash value prefixed by
$00\cdots$, then any element has probability $1/4$ of being in the cache
(and thus the cache will contain 25\% of all elements, unless it
overflows). More formally, the algorithm can be described as in
Figure~\ref{alg:as}.

\begin{figure}[h]\centering
%\begin{center}\vspace{1em}
  \begin{minipage}{6.4cm}\parindent0pt
      %\emph{Parameter:} ${m}$ cache capacity\\
      %\emph{Input:} a stream $\mathcal{S}=(s_1, \ldots, s_N)$\\
      %\hrule\vspace{.4cm}
      
      \textbf{initialize} $C := \emptyset$ (cache) and $d := 0$ (depth)

      \begin{tabbing}
        mm \= mm \= mmmmmmmm \= mm \= mm \= \kill
        \textbf{forall} $x \in \mathcal{S}$ \textbf{do}\\[0.5em]
        mm \= mm \= mmmmmmmm \= mm \= mm \= \kill
        \> \textbf{if} {$h(x) = 0^d\cdots$ and $x\not\in C$} \textbf{then}\\
        \> \> $C := C \cup \left\{ x \right\}$\\[0.5em]
        \> \textbf{while} $|C| > {m}$ \textbf{do}\\
        \>\> $d := d + 1$\\
        \>\> $C := \left\{x \in C\ |\ h(x) = 0^d\cdots\right\}$\\[0.5em]
        \textbf{return} $2^d \cdot |C|$
      \end{tabbing}
    \end{minipage}%\vspace{1em}
    \caption{\label{alg:as}The {\algAS} algorithm.}
%\end{center}
\end{figure}

%\noindent
In the end, the algorithm has a cache $C$ containing any element with
probability $\cramped{1/2^d}$; a good statistical guess of the entire
number of elements is the $\cramped{2^d}\cdot |C|$. This is what Flajolet
proved in his paper~\cite{PF084}, along with the accuracy of this
estimator.

\subsection{The wheels are greased: or how the analysis holds no
  surprises.}

In the context of Flajolet's papers on streaming algorithms, this paper is
interesting not for its complexity, but for its simplicity. Indeed, the
mathematical structure of the algorithm is, in essence, practically the
exact same as that of {\algAC} and {\algPC}. But the analysis is here much
clearer and simpler---it is only three pages long! This owes to the fact
that it is formulated in terms of splitting process~\cite{PF034}, and
benefits from the progressive refinement and simplification of that type
of analysis\footnote{Interestingly, as mentioned in
  Subsection~\ref{subs:nomathnoalg}, this method was also later used by
  Kirschenhofer {\etal}~\cite{KiPrSz92} to provide a simpler analysis of
  {\algPC}.}.

A splitting process simply means that we consider the execution of the
algorithm as a branching structure: a tree which contains at its root all
elements, and at each node separates the elements which are discarded (in
the left subtree) and those that are retained (in the right subtree); this
yields a functional divide-and-conquer type equation that has now become
easy to solve.

In the same vein, another contemporary article by
Greenberg~\etal~\cite{PF068}, on estimating the number of conflicts in
communication channels, bears more than passing ressemblance to this
algorithm and its analysis.

Other concepts, such as the previously oft-used ``\emph{exponential
  approximation}'', are now much better understood, routinely used in
fact, and no longer justified. In fact, this article marks the first time
Flajolet explicitly~\cite[\S 3.C]{PF084} states that the approximation
$(1-a)^n\approx \smexp{-ax}$ is equivalent to a \emph{Poissonization}: in
the splitting process, instead of considering all possible ways to split
$n$ values into two subtrees, Poisson variables of mean $n/2$ are
used---which yields a very precise approximation in practice\footnote{It
  would take several years for the reverse notion to appear: called
  \emph{Depoissonization}, it formalizes how to go from the Poisson model
  to the exact/Bernoulli model.}.

%%%\paragraph{The algorithm seen as a trie.}
%%%Finally, it is perhaps interesting to note that there is related
%%%combinatorial interpretation of the algorithm. Consider a \emph{binary
%%%  trie}, containing $n$ distinct infinite random $\set{0,1}$ strings (that
%%%is to say that each letter of a string is either $0$ or $1$ with
%%%probability $1/2$). Because the strings are uniform, you expect each node
%%%of the trie to evenly split the set of remaining strings---half to the
%%%left, and half to the right---and by induction, you expect all nodes of
%%%the trie that are at the same depth to contain roughly the same amount of
%%%strings. Thus, if you notice that some node at depth $d$ contains $x$
%%%strings, you can reasonably expect the whole trie to contain $x$ (the
%%%number of strings rooted at that node) times $\cramped{2^d}$ (the number
%%%of nodes at depth $d$) strings in total.

\subsection{As a sampling algorithm}

Despite conceptual strengths, {\algAS} is less accurate than {\algPC}, and
though implemented~\cite{AsScWh87} was, as far as I know, never used in
practice as a cardinality estimation algorithm. But Flajolet quickly
realized that it could be used to yield very interesting statistics beyond
the number of distinct elements it was initially designed to
estimate~\cite{PF134}.

Indeed, at any point during the course of its execution, the algorithm
(parameterized to use $m$ words of memory) stores a uniform sample of
between $m/2$ and $m$ distinct elements taken from the set underlying the
stream. That is to say elements are sampled independently of their
frequency in the stream: an element appearing a thousand times, and
another appearing only once would be sampled with equal probability.

Furthermore by attaching frequency counters to the elements,
the proportion of various classes of elements can be estimated: for
instance, those elements appearing once (called \emph{mice}, in network
analysis) or those appearing more than say, ten times (called
\emph{elephants}), see \cite{Louchard97, LoLuSw13} for detailed analyses.

%% TODO : mention impopularity

This algorithm was subsequently rediscovered by several authors, but in
particular by Gibbons~\cite{Gibbons01}, who most pertinently renamed it
\emph{Distinct Sampling}---which then influenced an algorithm by
Bar-Yossef \textit{et al.}~\cite[\S 4]{BaJaKuSi02}.

More recently, the basic idea was popularly generalized as
$\ell_p$-sampling, see for instance~\cite{MoWo10}, which samples an
element $i\in\set{1,\ldots, n}$, appearing $f_i$ times in the stream, with
probability proportional to ${f_i}^p$ for some specified $p\in\Rpos$---in
this setting, \emph{Distinct Sampling} would be related to the special
case $p=0$.

In another direction, Helmi {\etal}~\cite{HeLuMaVi12} have begun
investigating algorithms in the vein of \emph{Distinct Sampling}, but with
the novel feature of being able to control the size of the cache as a
function of the number of distinct elements (for instance, you may ask for
a uniform sample of $k\log n$ distinct elements).

\section{Epilogue}

The novel ideas behind these algorithms, and behind {\algPC} in
particular, had a lasting impact and contributed to the birth of streaming
algorithms. The concepts were further formalized in the groundbreaking
paper by Alon {\etal}~\cite{AlMaSz96} in 1996/2000, and from then on, the
literature, until then fledgling and rooted in practical considerations,
became increasingly expansive and theoretical.

Flajolet's own further contribution, the {\algLL} family of algorithms, is
generally much better known than its predecessors. These algorithms bring
small but crucial optimizations: a different statistic that requires a
logarithmic-order less memory to track~\cite{PF176}\footnote{This idea was
  first mentioned in the last few pages of Flajolet and
  Martin's~\cite{PF050} article; but at the time it was not apparent that
  the accuracy tradeoff was worth the gain in space---something later
  highlighted by Alon {\etal}~\cite{AlMaSz96}.}; some algorithmic
engineering to avoid extremal values and increase accuracy~\cite[\S
5]{PF176}; and the same gain in accuracy without algorithmic engineering,
but through a different averaging scheme involving the harmonic
mean~\cite{PF193}.

Although these evolutions might seem self-evident now, they also
considerably complexify the analysis of the algorithms: the math involved
in the analysis of {\algHLL} is severely more complex than that of
{\algPC}.

In the 2010s, with the continuing emergence and ubiquity of Big Data, the
{\algHLL} algorithm is universally recognized as the most efficient
algorithm in practice for cardinality estimation, and it is used by
influential companies~\cite{HeNuHa13}.

\section*{Acknowledgments}

I wish to extend my heartfelt gratitude to Nigel Martin: his kindness and
his willingness to share allowed me to unearth forgotten treasures.

For their help in filling in historical details, I would also like to
thank: Mikl{\'{o}}s Cs{\H{u}}r{\"o}s, Doug McIlroy, Mark Wegman,
Fr\'{e}d\'{e}ric Meunier, Marianne Durand, Piotr Indyk, Steven Finch,
\'{E}ric Fusy, Lucas Gerin, Pranav Kashyap, Mike Paterson.

Finally, I am deeply thankful to Brigitte Vall\'{e}e for her help, to
Mich\`{e}le Soria for her unwavering---and much welcome---support, and
Bruno Salvy for his help proofing this manuscript.

\nopagebreak

\bibliographystyle{plain}
\bibliography{misc,PhilippeFlajoletBibliographyAll,probcounting}

\end{document}